# Robust and Active Visible-Light Integrated Photonics on Thin-Film Lithium Tantalate for Underwater Optical Wireless Communications

**Changjian Guo,[1,3,†,*] Xingjie Li,[1,†] Xiaofeng Wu,[1,†] Jiajie Deng,[2] Jiayu Huang,[1] Yunjie Wang,[1] Wenchang Yang,[1] Weilong Ma,[1] Ziliang Ruan,[2] Kaixuan Chen,[1,3,*] Sailing He,[2] and Liu Liu[2,4,*]**

[1]Guangdong Provincial Key Laboratory of Optical Information Materials and Technology, South China Academy of Advanced Optoelectronics, Sci. Bldg. No. 5, South China Normal University, Higher-Education Mega-Center, Guangzhou 510006, China

[2]State Key Laboratory of Extreme Photonics and Instrumentation, College of Optical Science and Engineering, International Research Center for Advanced Photonics, East Bldg. No. 5, Zijingang Campus, Zhejiang University, Hangzhou 310058, China

[3]National Center for International Research on Green Optoelectronics, South China Normal University, Guangzhou 510006, China

[4]Jiaxing Key Laboratory of Photonic Sensing & Intelligent Imaging, Intelligent Optics & Photonics Research Center, Jiaxing Research Institute, Zhejiang University, Jiaxing 314000, China

[†]These authors contributed equally to this work.

**Corresponding author**

Changjian Guo (guochangjian@m.scnu.edu.cn), Kaixuan Chen (chenkaixuan@m.scnu.edu.cn), and Liu Liu (liuliuopt@zju.edu.cn).

**Abstract**

Visible-light integrated photonics enables compact platforms for sensing, precision metrology, and free-space data links at visible wavelengths. However, many applications remain limited by the lack of high-speed and robust modulators in the blue-green band. Here we report, both operating at 532 nm, thin-film lithium tantalate waveguides of propagation losses of dB/cm scale and modulators with a flat frequency response to ~50 GHz. The modulator remains stable when delivering 5 dBm modulated optical power for an hour, which cannot be achieved by thin-film lithium niobate based counterparts under similar conditions and structures. System-level underwater optical wireless communication (UWOC) is validated with 112-Gb/s transmission over 3-m and 64-Gb/s transmission over 9-m underwater links. This represents the first integrated external modulator based UWOC system, overcoming the bandwidth-power-chirp trade-offs of traditional directly modulated laser-based systems. We further demonstrate dual-drive modulators for optical single-sideband and electro-optic frequency-comb generations in the green-wavelength band. These results provide a foundation for complex, robust, and active visible-light photonic integrated circuits for underwater optical applications.

## Introduction

Visible integrated photonics plays a crucial role in various cutting-edge applications, including sensing, quantum information processing, precision metrology, augmented and virtual reality displays, and free-space visible light communications (VLC) [1-6]. These applications typically require photonic devices that offer high-speed modulation, high optical power handling, and long-term stability at visible wavelengths. For instance, underwater wireless optical communication (UWOC) stands out due to its ability to provide high data rates and low latency, offering significant advantages over traditional radio-frequency and acoustic methods in underwater environments. A UWOC system typically operates in the blue–green band, which coincides with the seawater's minimal-attenuation window. Applications like underwater surveillance, autonomous underwater vehicles, and marine research require real-time data transmission over a distance in free space, making high-performance light modulation a critical function in these systems [6]. Currently, UWOC systems primarily rely on directly modulated light sources such as laser diodes (LDs) and light-emitting diodes (LEDs), which operate within gigahertz modulation bandwidths [4]. Moreover, direct modulation introduces frequency chirp that broadens the optical spectrum and distorts signals, especially at high drive powers required for UWOC. These inherent limitations impede further scaling towards high-data-rate communications. As such, the development of high-bandwidth, high-power handling, and stable visible-light modulators, as well as photonic integrated circuits (PICs) at visible wavelengths, is essential for improving the performance of UWOC and related systems, and enabling more demanding applications.

Simultaneously, the broader field of visible integrated photonics is advancing rapidly, driven by PICs that support complex optical functions on a single chip. One recent work has established a complementary metal-oxide-semiconductor compatible Ge-doped silica photonics platform that provides a fiber-like, ultralow-loss passive backbone from the visible to the telecom band, with 0.32 dB/m waveguide propagation loss at 532 nm and micro-resonators with intrinsic Q factors exceeding 108 [7]. Meanwhile, significant progress has been made with micro-resonator based soliton microcombs, which provide phase-coherent multi-wavelength carriers and frequency references on a chip. These technologies, along with dispersion engineering and wavelength conversion strategies, are pushing the limits of integration into the blue–green wavelengths, further enhancing the capabilities of high-speed modulation and frequency generation [8-10]. Moreover, integrated optical beam steering offer solutions for target alignment and tracking, a critical function for maintaining high-quality communication links in dynamic environments like free-space optical and UWOC systems, as well as light detecting and ranging (LiDAR) applications [11-12].

Despite these advancements, electro-optic (EO) modulators at visible wavelengths still face significant challenges in meeting the demands for high bandwidth, low drive voltage, low insertion loss, and high optical power robustness. Platforms such as silicon nitride or silica as mentioned above can provide ultralow-loss waveguides and circuits, but they do not natively possess any high-speed EO properties, and therefore typically require heterogeneous integration or extra chips for fast signal modulation [13]. Thin-film lithium niobate (TFLN) modulators have been adopted recently in optical systems at telecom wavelengths to meet the above modulation requirements [14-15]. Extending TFLN modulators to visible wavelengths is feasible. Demonstrations about 532-nm modulators with modulation bandwidths beyond 25 GHz [16], 738-nm modulators with voltage-length product ($V_\pi \cdot L$) of 0.55 V·cm [17], and 450-nm modulators with $V_\pi \cdot L$ as low as 0.17 V·cm [18] have been reported. Yet, no real data transmission has been achieved on the TFLN platform. Challenges such as photorefraction and optical damage largely influence the performance of TFLN-based modulators at shorter wavelengths or higher optical powers. Photorefractive charge transport and defect-mediated dynamics in lithium niobate materials can lead to bias drift and transient power or phase fluctuations as optical power increases [19]. Even permanent photodamages can happen at higher optical intensities due to the relatively low optical damage threshold (in the order of $10^3$ W/cm² at green wavelengths) for lithium niobate. This phenomenon becomes more severe on TFLN, as the light intensity is largely increased as compared to the case with a bulk crystal [20]. This presents a limitation for many visible-light applications, including UWOC systems, which require high optical power to overcome water attenuation and turbulence.

Thin-film lithium tantalate (TFLT) offers a potential solution to this problem. While TFLT shares similar EO properties with TFLN ($\gamma_{33} \approx 30$ pm/V for TFLT), it provides a much higher optical damage threshold and therefore reduced susceptibility to photorefraction, making it a more robust material for high-power visible-light PICs [21-24]. Prior studies have shown that TFLT-based resonators exhibit negligible photorefraction or drift under a high optical power at infrared wavelengths [21, 22]. Furthermore, TFLT modulators have been demonstrated to handle high optical powers, maintaining performance stability even at elevated power levels [23]. Recent researches have also demonstrated the successful integration of various TFLT-based devices into a circuit, highlighting their scalability for volume manufacturing [24].

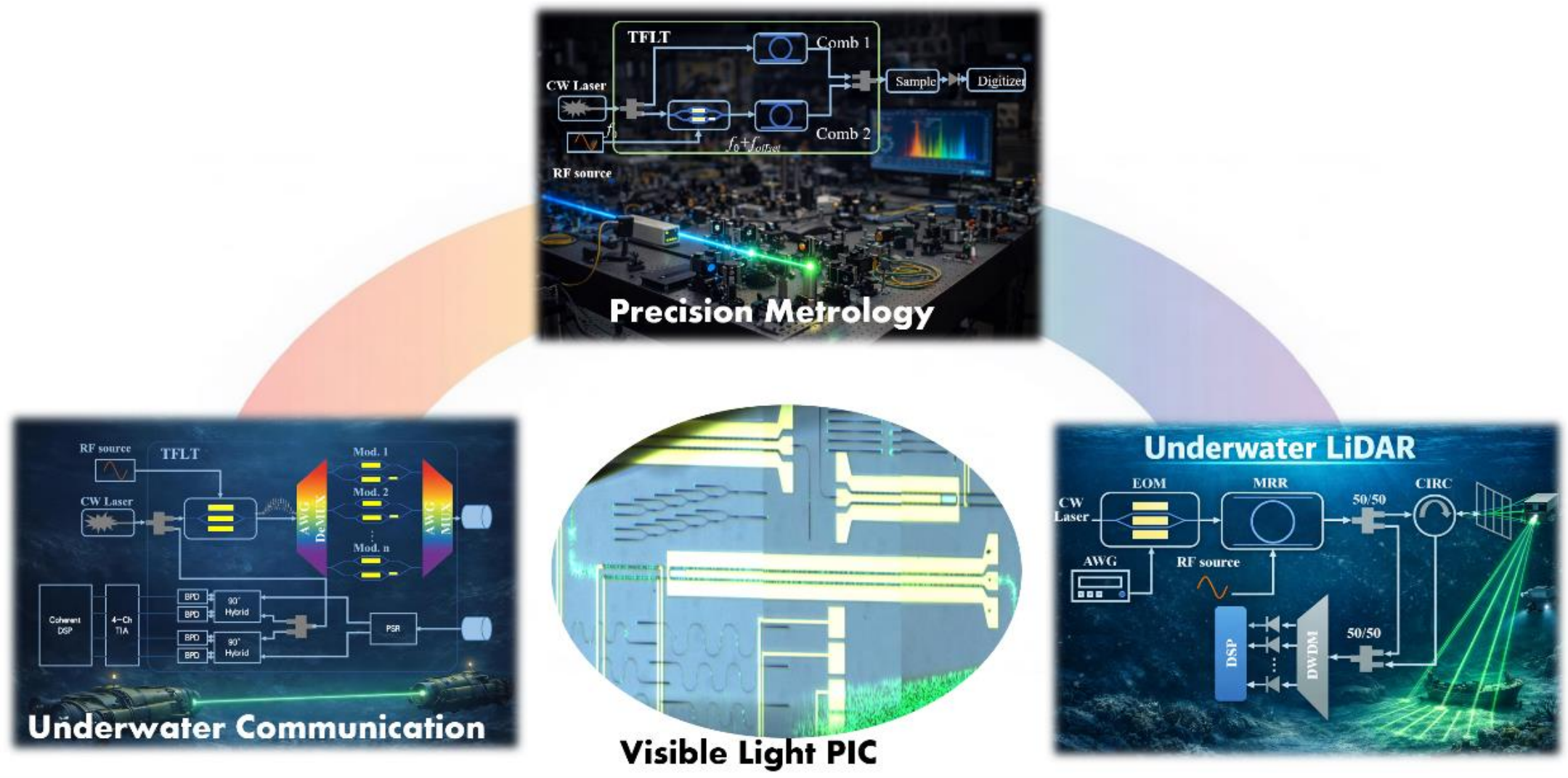

**Figure 1 | Applications of the visible light PICs.**

In this paper, we report a high-performance PIC based on TFLT, tailored for UWOC applications, including integrated Mach-Zehnder modulators (MZMs) and EO frequency comb devices operational at 532 nm. These devices feature low optical loss waveguides and efficient fiber-to-chip grating couplers (GCs) to ensure a high optical power throughput. We demonstrate that the fabricated TFLT modulator can handle continuous-wave on-chip optical power of ~11.5 dBm without any degradation in performance. The measured half-wave voltage reaches 2.6 V for a 3-mm long modulation section, and the 3-dB modulation bandwidth exceeds 50 GHz. Using this modulator, we successfully demonstrate the transmission of 100-Gb/s on-off key (OOK) and 56-Gbaud four-level pulse amplitude modulation (PAM-4) signals over a 3-m underwater link, as well as 64-Gb/s OOK signals over a 9-m underwater link. A Dual-drive- (DD-) MZM for optical single-sideband and EO frequency-comb generations in the green-wavelength band is also demonstrated. To our knowledge, this is the first integrated TFLT modulator and frequency comb devices operating in the blue-green spectrum, as well as the first UWOC system using the external modulation approach. The combination of robustness and active functionalities in the present TFLT platform also makes it outperform other photonic integration platforms, where high-performance modulation is absent, for applications at visible wavelengths as shown Fig. 1.

**Device design**

The modulator is implemented as a traveling-wave MZM with a push-pull configuration and an on-chip 50Ω termination resistor. The device structures and some images of the fabricated devices is illustrated in Fig. 2. Light is first coupled into the chip through a surface GC [25], which is optimized for a Gaussian input beam input with a 3.5-μm waist diameter by adjusting the duty cycle ($\rho$) and period ($p$). After in-coupling, the light propagates through an adiabatic inverse taper, which narrows the waveguide core from 5 μm to 0.4 μm. This taper is designed to suppress higher-order modes, ensuring an efficient excitation of the fundamental transverse-electric ($TE_0$) mode throughout the device. We refer to Supplemental Note 1 for detailed mode properties of the present TFLT ridge waveguide. 90° Euler bends with a waveguide width of 0.4 μm and an equivalent bend radius of 70 μm is also employed as a filter for higher-order modes. Following these structures, a multimode interferometer (MMI) is used as the 3-dB splitter and combiner of the MZM. In the modulation section, the waveguide is adiabatically widened to 1.5 μm. This significantly reduces scattering losses, as discussed in the next section, and enhances the field overlap between the optical and microwave fields, and thus, improves the modulator's overall efficiency.

The co-planar waveguide-based capacitively loaded travelling-wave (CLTW) electrodes are utilized to ensure broadband operation. The CLTW electrodes are made of gold with a thickness of 1.1 μm to minimize microwave losses. The width of the uncoupled signal (S) electrode ($w_s$) is 40 μm, while the width of the ground (G) electrode ($w_g$) is set to 80 μm. The remaining parameters of the CLTW electrodes are optimized to achieve impedance and refractive index matching, with ($g$, $r$, $c$, $s$, $t$, $h$) = (3.5, 45, 5, 2, 2, 3.5) μm, respectively [16, 26]. The simulated performances and EO modulation frequency response of the CLTW electrodes are shown in Fig. 3. From the simulation results, it can be observed that the designed electrodes can achieve an excellent refractive index and impedance matching, which benefits from the large group refractive index of the $TE_0$ mode of the present TFLT waveguide at the visible-light band. Moreover, due to the large size of the signal electrode, even when the microwave frequency reaches 100 GHz, the microwave loss remains at a low level of approximately 5.5 dB/cm. As shown in Fig. 3d, the simulated 3-dB bandwidth of the electrode can well exceed 120 GHz, far beyond that of the available photodetectors (PDs) at 532 nm wavelength.

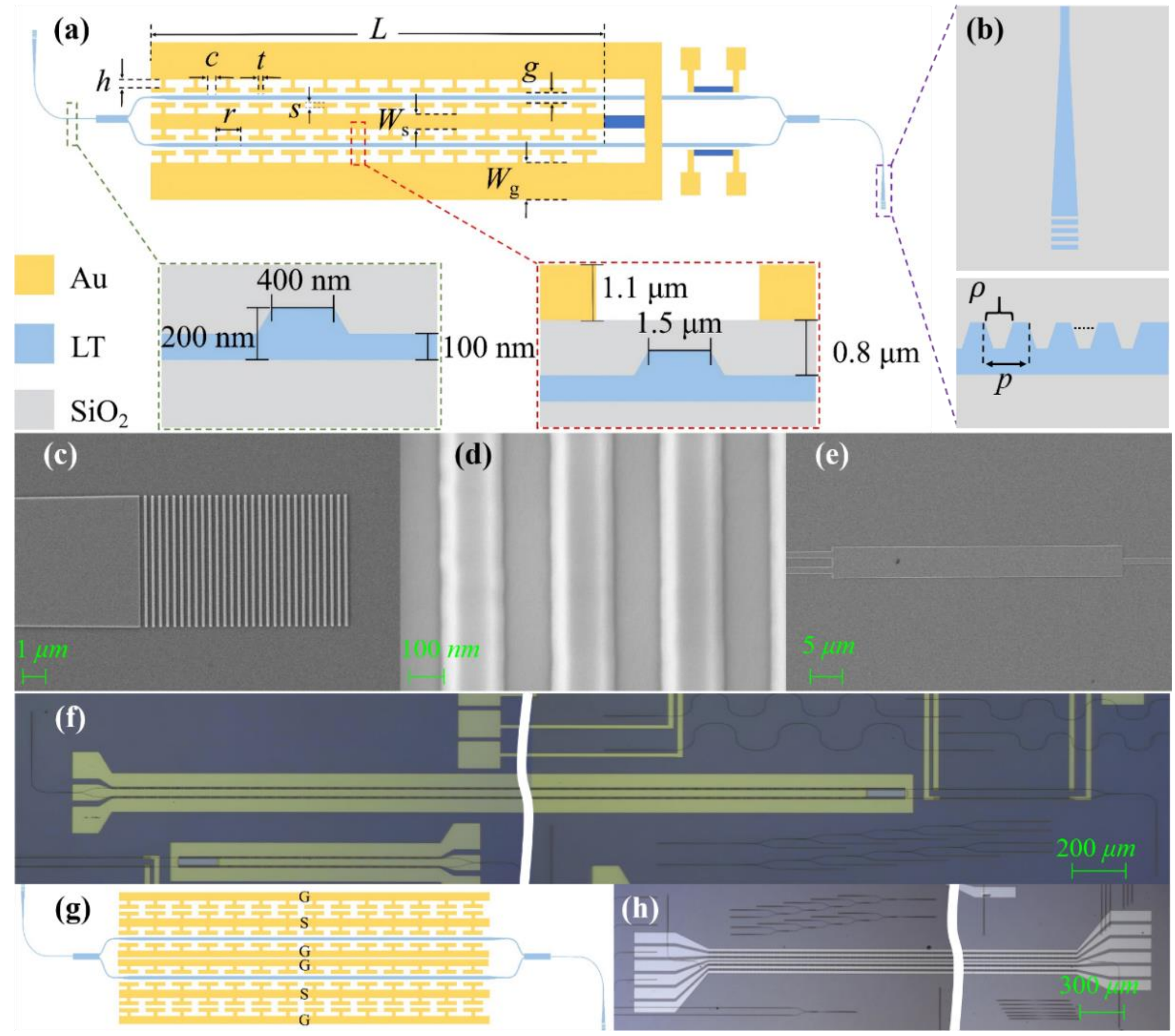

**Figure 2 | TFLT passive and modulator structures. (a)** Schematic diagram of the TFLT MZM, showing the input/output GCs, the MMI splitters, and the traveling-wave electrodes. **(b)** Detailed sketch of the GC. **(c)-(e)** SEM images of the GC, the grating lines, and the MMI, respectively. **(f)** Microscope image of the fabricated intensity modulator chip. **(g)** Schematic diagram of the DD-MZM on TFLT with a GSGGSG electrode configuration. **(h)** Microscope image of the fabricated DD-MZM.

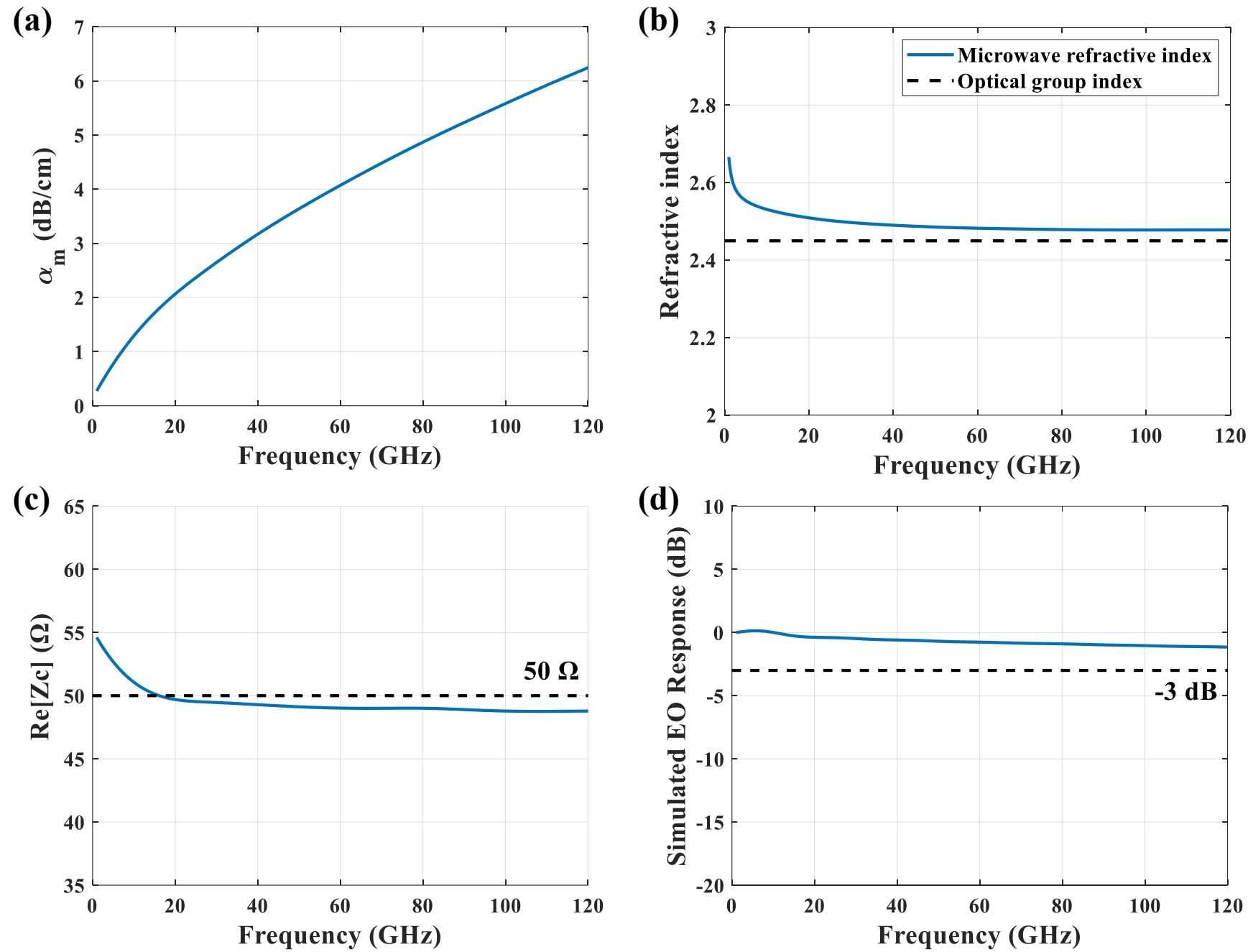

**Figure 3 | Simulated modulator performance for 532 nm.** The simulated **(a)** microwave loss $\alpha_m$, **(b)** microwave refractive index $n_m$, **(c)** characteristic impedance $Z_c$, and **(d)** EO modulation frequency response of the CLTW electrodes.

## Passive component characterization

We first characterized the fiber–chip coupling loss using a supercontinuum source and GCs designed for 532 nm. At the optimized parameters of $p$ = 285 nm and $\rho$ = 0.7, a maximal simulated coupling efficiency of 45% (corresponding to a loss of 3.5 dB per coupler) is obtained at 532 nm. On the other hand, the measured insertion loss of a pair of GCs is ~9 dB at 532 nm, as both shown in Fig. 4a, corresponding to ~4.5 dB coupling loss per GC, slightly higher than the simulation result. The measured spectral response exhibits a ~17-nm 3-dB bandwidth. It is worthwhile to note that substantially higher coupling efficiency at 532 nm can be achieved by optimizing the layer structure of the TFLT wafer [25].

We then quantified the propagation loss of TFLT ridge waveguides at 532 nm for different core widths using a serial of spiral waveguides of different lengths (See Supplemental Note 2). As shown in Fig. 4b, the propagation loss decreases remarkably when the waveguide width increases from 0.4-μm to 1-μm. For 0.4-μm-wide waveguides, the measured propagation loss reached around 4 dB/cm, while it reduces to around 1.5 dB/cm for 1-μm width. The propagation loss remains roughly the same when further increasing the waveguide width to 1.5 μm. Combined with the demonstrated low-loss GCs on the same platform, these high-performance passive devices outline a clear route toward low-loss and large-scale visible-light PICs on the TFLT platform.

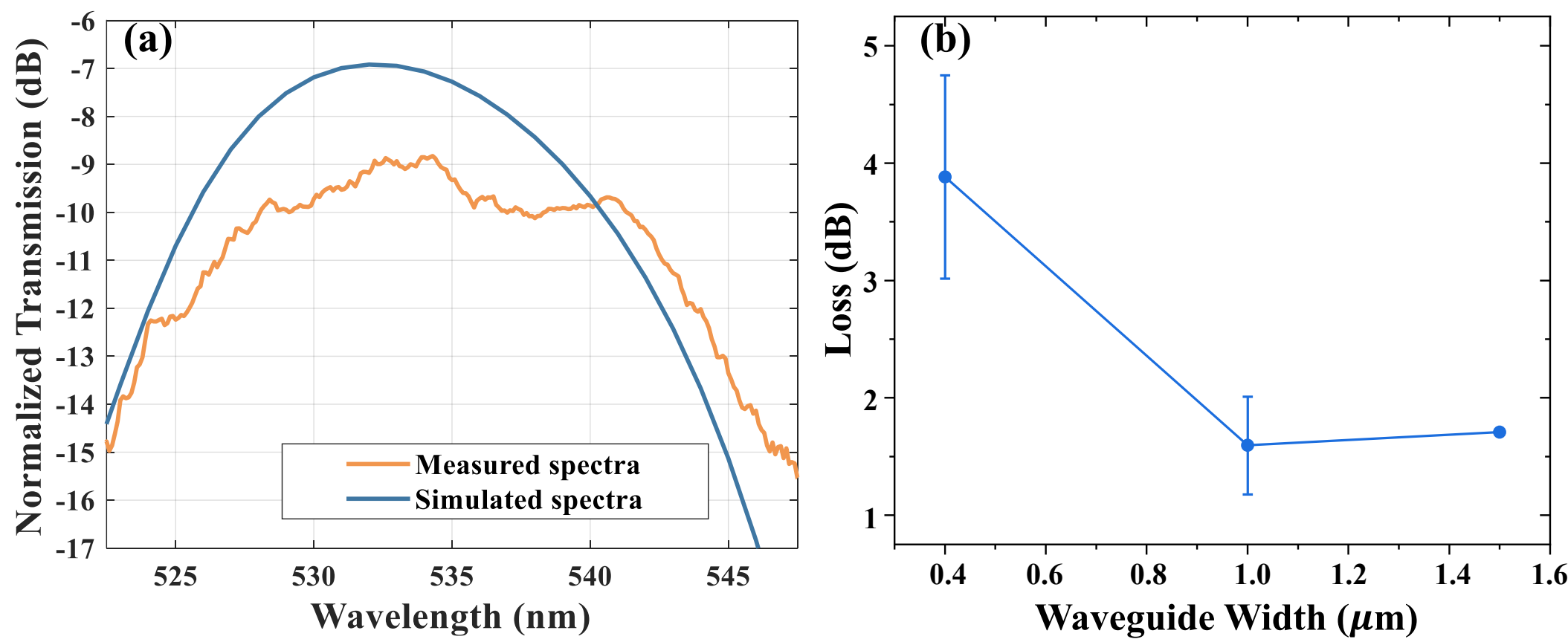

**Figure 4 | Measured performances of passive components. (a)** Measured and simulated back-to-back loss of a pair of GCs. **(b)** Measured TFLT waveguide loss as a function of the waveguide width.

### Modulator characterization

We characterized the TFLT modulator under both steady-state and high-speed driving conditions to evaluate its robustness for VLCs. The total fiber-to-fiber insertion loss of a single modulator at 532 nm was measured to be approximately 12 dB. After subtracting the coupling loss from the input and output GCs, the on-chip insertion loss of the modulator would be 2.7 dB.

The half-wave voltage ($V_\pi$) of the modulator was measured using a 100-kHz sawtooth voltage sweep. The normalized transmission as a function of bias voltage shows a clear sinusoidal transfer function in Fig. 5b, yielding $V_\pi \approx 2.6$ V. Considering the 3-mm modulation length, this corresponds to $V_\pi \cdot L \approx 0.78$ V·cm, in excellent agreement with the simulated value of 0.77 V·cm. Note that the $V_\pi \cdot L$ value can be further reduced as decreasing the gap between the electrodes [26]. The modulator is also equipped with thermo-optical (TO) phase shifters of lengths of 0.51 mm for low-speed bias tuning. As shown in Fig. 5c, a $P_\pi$ of 50 mW can be measured for the TO phase shifter. One can also extract the static extinction ratio (ER) here, which reaches >35 dB, indicating excellent power balances in the MMI splitters and two interferent arms. Next, we evaluated the high-frequency modulation performance using the setup shown in Fig. 4a. As seen in Fig. 5d, the measured small-signal EO response remains flat up to approximately 50 GHz. The slight roll-off and fluctuations beyond 50 GHz can be attributed to the bandwidth limitation of the visible-light PD. Thanks to the CLTW electrode employed here, >100 GHz 3-dB bandwidth can be anticipated in the present modulator, which facilitates to support high data-rate transmission systems as demonstrated below.

We also examined the long-term power stability of the TFLT modulator at high input optical powers. A TFLN based modulator of a similar geometry was also fabricated and measured as a comparison. For both devices, the operating points at the maximal transmission were set using the TO phase shifter. As shown in Fig. 6a, when the input optical power exceeds 10 dBm, the TFLN modulator presents rapid power fluctuations caused possibly by photorefractive effects [19, 20]. In contrast, from Fig. 6b, the transmission of the TFLT modulator maintains stable at input powers up to 18 dBm for an hour, with output powers >5 dBm. This corresponds to a stable on-chip optical-power handling capability up to 11.5dBm for the present TFLT modulator. Crucially, this >5 dBm stable output power is achieved without requiring any auxiliary stabilization techniques (e.g., bias feedback circuits) that are often necessary for TFLN modulators due to photorefractive effects (Fig. 6a). This result confirms the robustness of the TFLT platform for high-power and short-wavelength EO modulations.

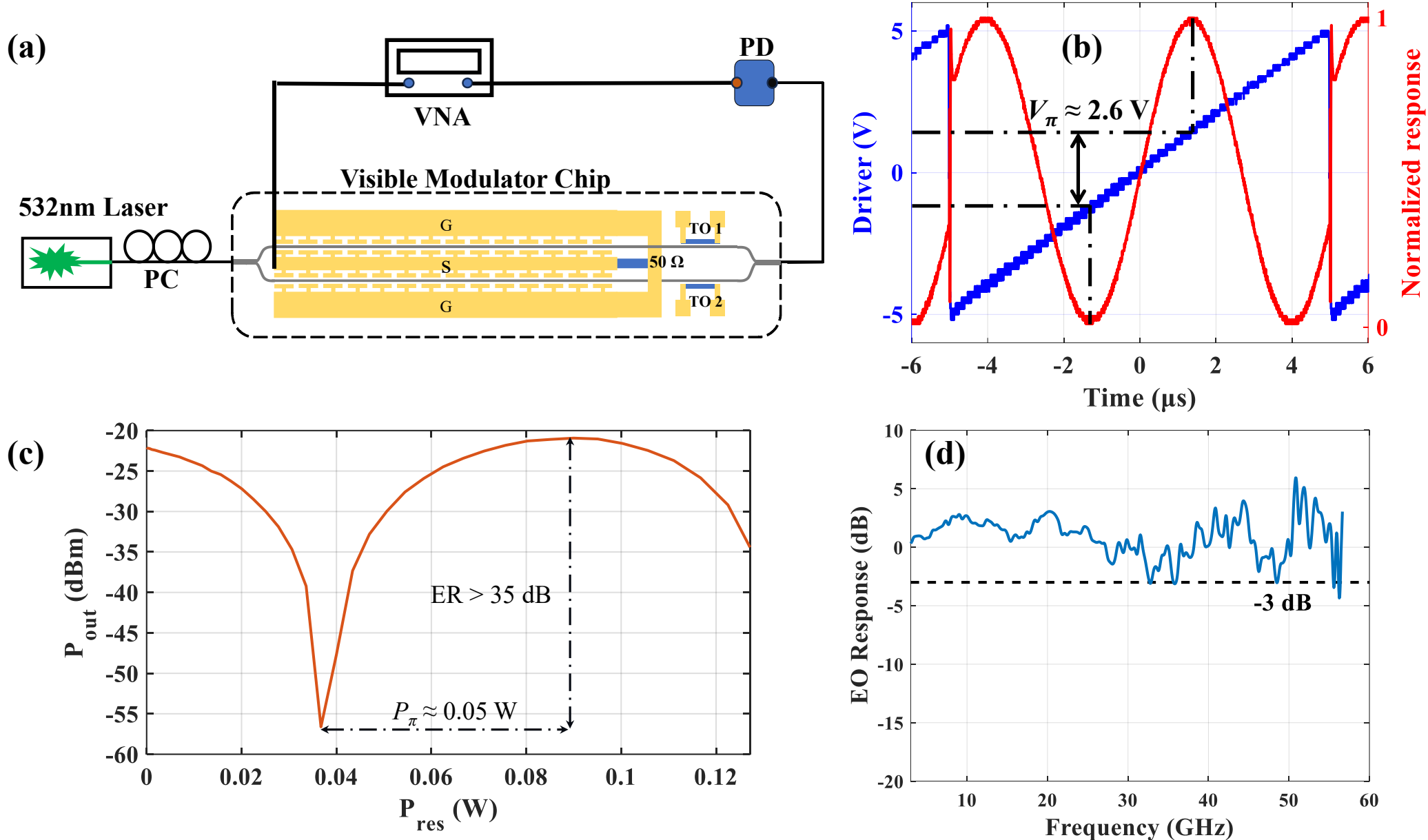

**Figure 5 | EO modulation and TO bias tuning performance. (a)** Setup for the modulator bandwidth measurement. **(b)** Normalized optical transmission as a function of the applied modulation voltage driven by a 100-kHz sawtooth signal. $V_\pi$ of 2.6 V can be extracted. **(c)** Transmission of the modulator as a function of the bias electrical power on the TO phase shift, showing an ER of ~32 dB. **(d)** Measured EO response of the modulator.

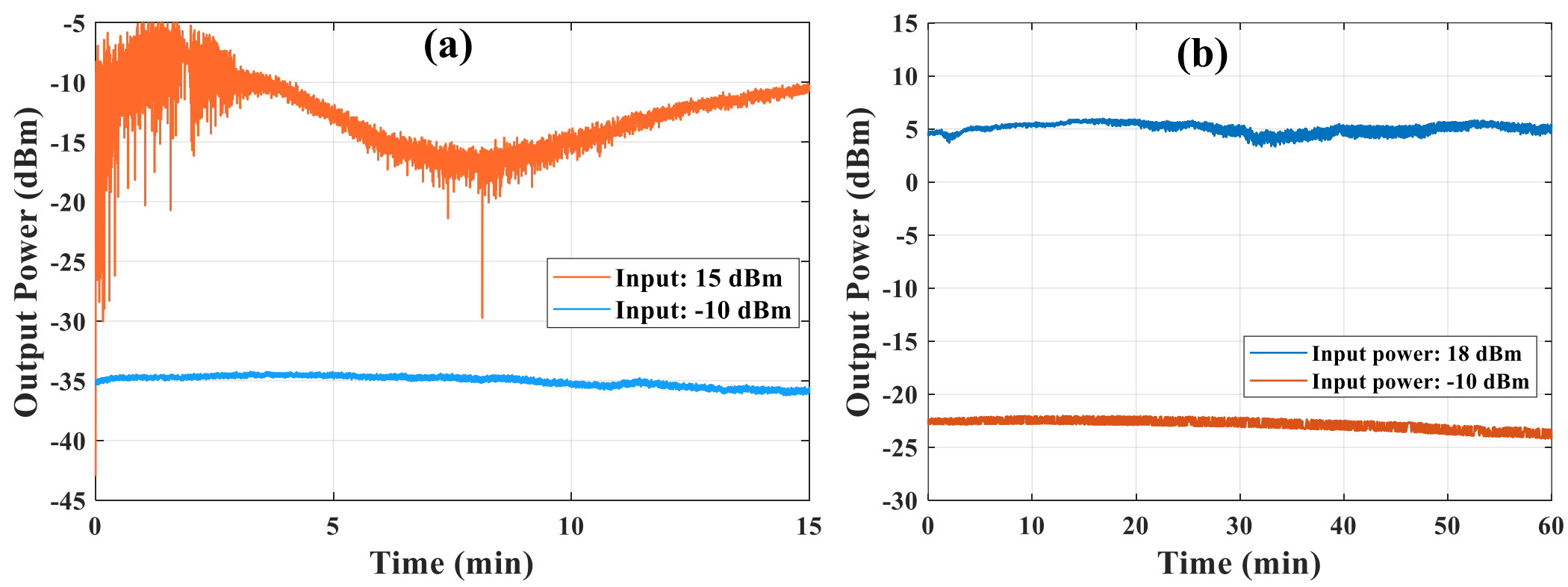

**Figure 6 | Stability performance of the modulators.** Temporal variation of the output optical power over time for **(a)** TFLN and **(b)** TFLT modulators under different input optical power levels.

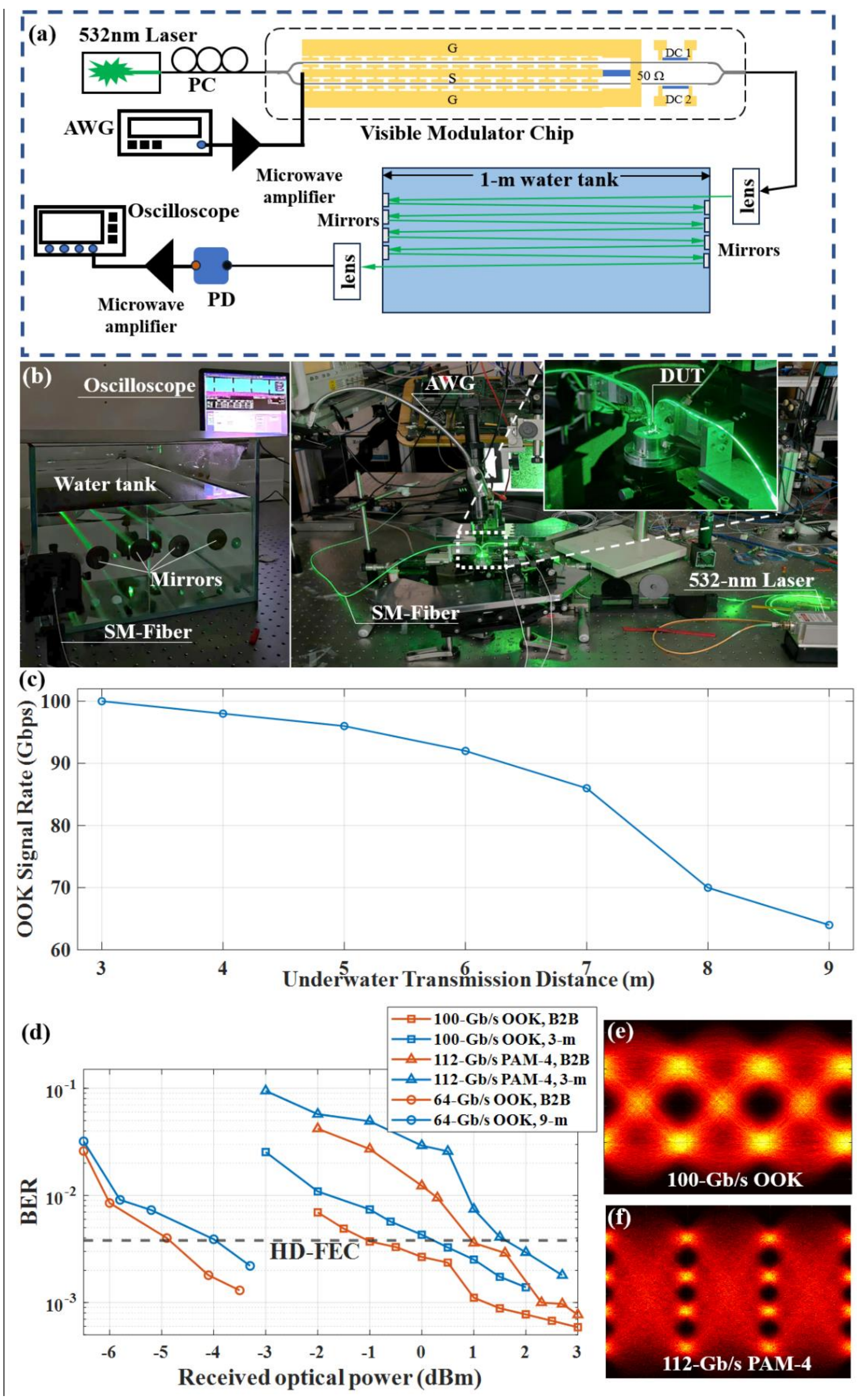

**Figure 7 | UWOC transmission performance.** (**a**) Schematic of the experimental setup for UWOC measurements. (**b**) Photos of the UWOC experimental setup. (**c**) Achievable data rate as a function of the transmission distance with OOK modulation. (**d**) Measured BER versus received optical power for the UWOC link. (**e**) Eye diagram of the 100 Gb/s OOK signal after transmission through a 3-m water channel using the fabricated TFLT modulator device. (**f**) Eye diagram of the 112 Gb/s PAM-4 signal under the same conditions.

### UWOC demonstration

To demonstrate the feasibility of the fabricated TFLT modulator in a real communication system, we further evaluated the performance of the device for high-speed UWOC applications using the setup sketched in Fig. 7a. A laboratory water tank of 1-meter length was filled with clear, particle-settled tap water, and up to 8 mirrors were placed inside to emulate a 9-meter underwater transmission link as shown in Fig. 7b. We first evaluated the achievable OOK data rate versus transmission distance. Using a bit error rate (BER) of $3.8 \times 10^{-3}$, i.e., the hard-decision forward error correction (FEC) threshold, as the criterion, 100-Gb/s OOK transmission was maintained for link lengths up to 3 m, limited by the maximum data rate available in the testing equipment. The achievable OOK rate below the BER threshold then decreased with distance, reaching 64 Gb/s over the 9-m path (Fig. 7c). BERs were thoroughly measured as a function of received optical power under back-to-back, 3-m, and 9-m conditions (Fig. 7d). At 3 m transmission distance, both 112-Gb/s PAM4 and 100-Gb/s OOK signals achieved BERs below the FEC threshold, while 64-Gb/s OOK transmission remained below the threshold over the 9-m path. The received-power penalties relative to the corresponding back-to-back cases were approximately 1 dB for both the 3-m and 9-m links. Although reported attenuation coefficients at 532 nm are only about 0.38 dB/m for tap water and 0.25 dB/m for deep-sea water [27], [28], the folded 9-m link exhibited a total loss of about 8 dB here due to additional mirror reflection and alignment losses. Nevertheless, the UWOC demonstrations here prove that the present TFLT modulator can deliver high-speed signals through a real water channel—a critical step not yet achieved with integrated visible-light modulators. It is worthwhile to note that the data transmission performance shown here is still limited by bandwidths of the equipment at both the transmitter and receiver sides, as well as the PD, rather than the modulator chip.

### Optical single side-band (OSSB) and optical frequency comb (OFC) generation

Besides a single modulator, the present TFLT platform also enables integration of more complicated optical and electrical structures to support advanced PICs. Here, we also demonstrate OSSB modulation and OFC generation at 532-nm using a dual-drive MZM (DD-MZM), and the measurement setup is shown in Fig. 8a. In this device, the two arms of the MZM are driven by separate CLTW electrodes. By tuning the microwave phase imbalance and the modulator bias, the sideband suppression ratio (SSR) and the optical carrier-to-sideband ratio (OCSR) can be programmed, enabling flexible complex-field synthesis for heterodyne detection and frequency-offset generation (see Supplemental Note 3 for detailed analyses). From Fig. 8b, one can see that an SSR larger than 15 dB can be obtained. Considering the lacking of a good optical signal amplifier in the visible wavelengths, this OSSB generation could be a key technique in future coherent VLC systems to improve the receiving sensitivity and the spectral efficiency.

Leveraging the same DD-MZM chip together with a high-power microwave amplifier, we also demonstrated OFC generation at 532-nm, obtaining 11 clearly resolved frequency lines with 50-GHz frequency spacing, as shown in Fig. 8c. Among this comb spectrum, 7 flat-top lines are obtained, which is achieved by optimizing the MZM bias and the inter-arm microwave phase and amplitude imbalances (see Supplemental Note 3 for detailed analyses) [29]. This OFC device can be employed as a laser bank at visible wavelengths. Combining with the high-performance passive devices on the same TFLT platform, this also enables wavelength-parallel photonic signal transmission using wavelength-division-multiplexing (WDM) technology. The present comb span is primarily limited by the short modulation length (3 mm) and the limited microwave driving strength. This constraint however, is not fundamental. Increasing the modulation length or employing a high-power drive would substantially increase the number of visible comb lines, and hence, wavelength channels for communications. However, this is at the expense of decreasing the power of each single frequency and possibly the data rate of each wavelength channel. A good balance should be considered here while maximizing the aggregate date rate of the WDM system.

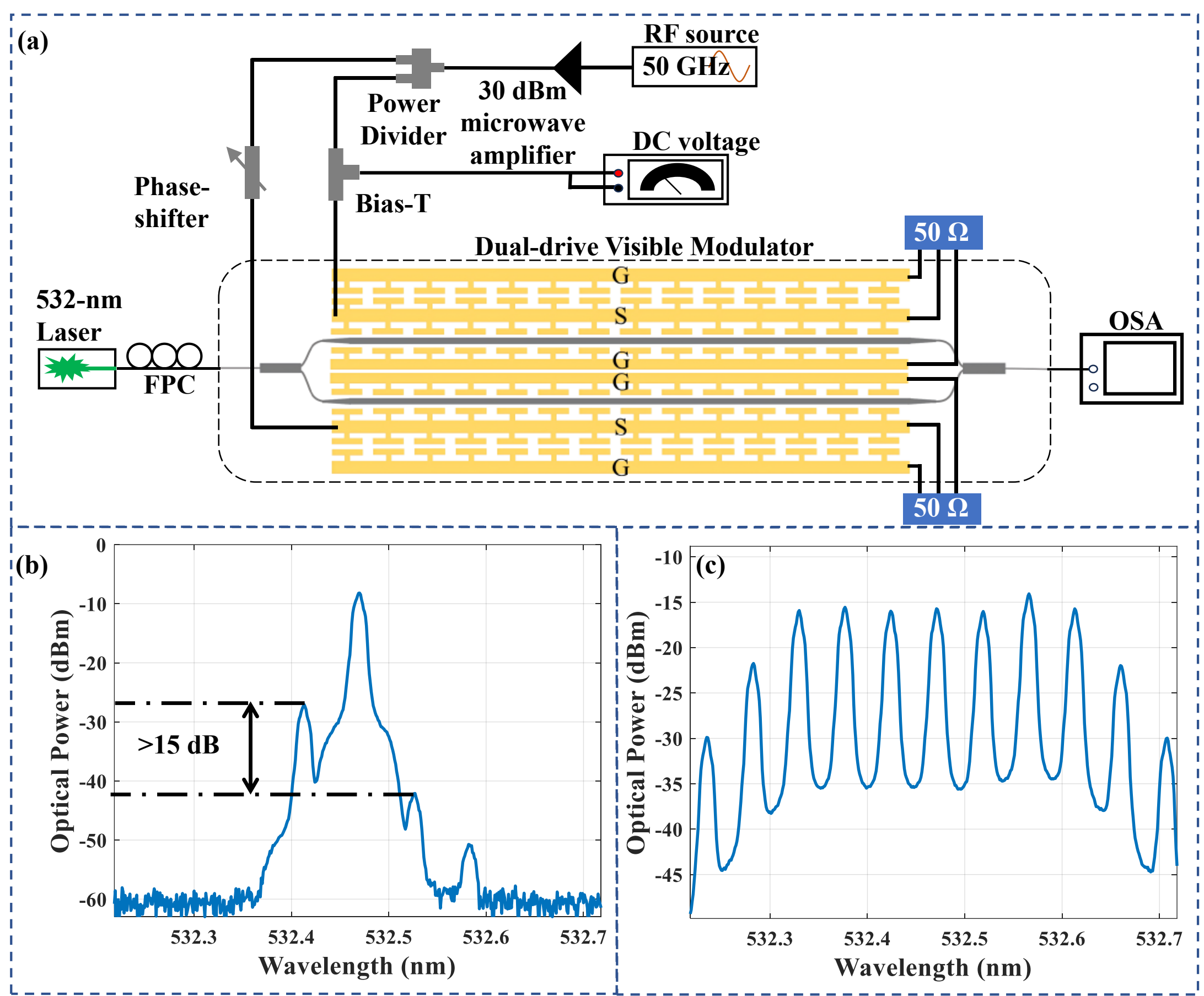

**Figure 8 | OSSB modulation and OFC generation. (a)** Schematic of the experimental setup for OSSB modulation and OFC generation. Measured optical spectra for **(b)** SSR and **(c)** OFC under a drive frequency of 50 GHz.

## Discussion and conclusion

We have demonstrated a TFLT EO modulator that operates at 532-nm wavelength. The key achievement of our work is combining the high modulation performance with a high optical power handling in a single integrated device. The achieved TFLT modulators exhibits $V_\pi{\cdot}L$ of ~0.78 V·cm and a modulation bandwidth of ~50 GHz. A high output optical power of 5 dBm is supported from the modulator chip with hour-scale stability. These performances enabled the demonstration of the real high-speed UWOC system. In Table 1, we compare the transmission performances using the present device with the recent UWOC/VLC systems using directly modulated LDs, injection-locked or feedback-assisted LDs, and high-speed LEDs or micro-LEDs [28-35]. With these sources, the achievable single-channel throughput in underwater links is typically in the range of 10–50 Gb/s. In our work, we have achieved 112-Gb/s transmission over a 3-m underwater link at 532 nm and 64-Gb/s transmission over a 9-m link. To the best of our knowledge, this is the first high-speed UWOC demonstration using an external electro-optic modulator. In contrast to directly modulated systems, which are fundamentally limited by laser chirp at high power and the intrinsic bandwidth-power trade-off, the present external modulation architecture using the present TFLT modulator eliminates chirp-induced distortion and allows the modulator's full >50 GHz bandwidth to be harnessed, enabling a single-channel data rate (112 Gb/s) that surpasses the physical limits of direct modulation.

**Table 1 | Performances of recent VLC/UWOC system demonstrations.**

| Ref. | Scenario | Wavelength | Tx type | Tx bandwidth (–3 dB) | Modulation format | Per-channel rate (Gb/s) |
|---|---|---|---|---|---|---|
| **This work** | UWOC | 532 nm | **TFLT MZM** | **~50 GHz** | PAM-4 | **112** |
| 2024 [28] | VLC | Visible (10λ) | 10λ LD | N/A | DMT | 12.61 max. |
| 2024 [29] | VLC | Visible (10λ) | 10λ LD | 3 GHz max. | DMT | 14.48 max. |
| 2022 [30] | VLC | RGB (3λ) | Tricolor LD | 4.06 GHz max. (-20 dB) | DMT | 17.168 max. |
| 2024 [31] | VLC | RGBP (4λ) | Integrated RGBP LEDs | 929 MHz max. | DMT | 6.86 max. |
| 2024 [32] | VLC | 451 nm | Blue LD | 5.9 GHz | DMT | 20.06 |
| 2020 [33] | UWOC | Visible (5λ) | 5λ WDM × 2 pol. | 18.46 GHz max. | PDM PAM-4 | 50 |
| 2017 [34] | UWOC | 450 nm | Blue GaN LD | 1.5 GHz | DMT | 12.4 |
| 2025 [35] | VLC | Visible (34λ) | 50 visible LDs | N/A | DMT | 15.66 max. |

Tx: transmission; DMT: digital multi-tone; PDM: polarization-division multiplexing.

The direct comparison with TFLN devices of a similar structure highlights the material advantage of TFLT against photorefraction-induced instability at blue-green wavelengths, which establishes TFLT as a practical platform for not only visible-light EO modulators but also robust visible-light PICs. The propagation loss of the present TFLT waveguides is still limited by the present fabrication processes. The etching induced sidewall roughness would large increase the propagation loss in the single-mode region. As the waveguide wide increases, the propagation loss drops. However, this phenomenon starts saturate beyond 1-μm width, as shown in Fig. 4b. Therefore, for a wider waveguide, the loss is likely determined by the quality of the oxide over-cladding layer prepared by an unoptimized deposition process. Through optimizing the fabrication, the propagation loss of a TFLT waveguide at visible wavelengths could likely be decreased to below 0.1 dB/cm as achieved in the silicon nitride platform with a similar index contrast [38]. Nevertheless, the propagation loss of 1-dB/cm-scale in the present chip can already support high performance passive circuits, such as wavelength filters using rings, gratings, or various interreference structures. Combined with the SSB and OFC generation functionalities demonstrated here, a single-chip WDM and coherent transmission system at visible wavelengths can be envisioned on this TFLT platform. These results represent a significant step towards next-generation ultra-fast and robust PICs for underwater communications and other demanding photonic systems using TFLT platform, as anticipated in Fig. 1.

## Methods:

**Device fabrication:** The device was fabricated on a commercially available blackened x-cut lithium-tantalate-on-insulator wafer, consisting of a 3-μm thick buried oxide layer and a 200-nm thick TFLT layer. The GCs, waveguides, and 3-dB MMI splitters were defined by electron-beam lithography and etched using inductively coupled plasma etching to form ridge waveguides that guide TE-polarized light. A ridge height of 100 nm, half the thickness of the TFLT layer, was selected to ensure single-mode guidance at 532 nm. The sidewall angle of the waveguide patterns is about 60° as determined by the etching process. The device was subsequently covered with an 800 nm thick $SiO_2$ layer, deposited by plasma-enhanced chemical vapor deposition. No any post annealing was adopted. In the final metallization step, coplanar CLTW electrodes made of gold, as well as the TO phase shifters and termination resistors made of titanium, were formed by thermal evaporation, and followed by a lift-off process. For the TFLN modulator for comparison, the same wafer stack and fabrication processes were employed, except that high temperature annealing was adopted after the waveguide patterning as described in [18] with anticipation of reducing the photorefractive effect.

**Dynamic measurement:** The experimental setups for EO bandwidth measurement, UWOC system, and SSB/OFC generation are shown in Figs. 5a, 7a and 8a, respectively. The electrical signal was fed to the device electrode through a 67-GHz-bandwidth microwave probe (GGB 67A). A 50-Ω on-chip termination was used for impedance matching. For the optical signal, a continuous-wave laser at 532 nm was coupled into and collected from the TFLT GCs using Nufern 460-HP single-mode fibers (operating wavelength range at 450–600 nm) with a 2.5-μm core diameter and a 3.5-μm mode-field diameter, after a manual polarization controller (PC). The modulated light from the output coupler of the device was collected by a high-speed PD working in visible wavelengths (Newport 1014). A vector network analyzer (VNA, Agilent PNA-X N5247A) was used for the bandwidth measurement, and the frequency response of the PD was deduced from the measured EO responses. For the UWOC experiments, an arbitrary waveform generator (AWG, Micram DAC1002) with a sampling rate of 100 Gsa/s was used to generate the electrical signals. A 50 GHz broadband amplifier (SHF 807c) with an output saturation peak-to-peak voltage of ~6 V was used to amplify the driving signal to the modulator. Bias of the MZM was tuned to the quadrature point with the TO phase shifter. For the BER measurements, the detected signal was sampled by a real-time oscilloscope (Tektronix DPO73304) and analyzed using off-line digital signal processing, including resampling, timing phase recovery, linear equalization with 31-tap feed-forward equalizer, 5-tap decision-feedback equalizer, and symbol decision. For the OSSB and OFC generations, an analog microwave source (Keysight 8257D) was used to generate the 50-GHz signal tone. A wide-band (20~50 GHz) and high-power (30 dBm) amplifier was used to amplify the driving signal. A microwave power splitter was used to divide the amplified electrical signal into two paths. One path passed through a tunable phase shifter, while the other path was routed through a bias-T. The two microwave outputs were then applied to the two input electrodes of the DD-MZM. The modulator output was connected to an optical spectrum analyzer (Yokogawa AQ6373E) to monitor the generated optical spectra. By jointly adjusting the relative phase and the bias of one arm through the bias-T, we optimized the OCSR in the SSB generation and comb spectral flatness in the OFC generation. All measurements were done with the chip mounted on a temperature-controlled stage at room temperature.

**Acknowledgements**

This work was supported by the National Natural Science Foundation of China (62435016, 62135012), the National Key Research and Development Program of China (2022YFB2804101), Basic and Applied Basic Research Foundation of Guangdong Province (2024A1515011710), Guangdong Provincial Key Laboratory of Optical Information Materials and Technology (2023B1212060065), and Science and Technology Planning Project of Guangdong Province (Grant No. 2019A050510039).

**Author contributions**

L.L. and C.G. developed the idea. C.G., K.C., and L.L. conceived modulator design. X.L., X.W., and J.D. performed the measurements for passive devices. X.L., X.W., and W.Y. performed the dynamic characterization and transmission experiment. X.L., X.W., W.M., and Z.R. developed the fabrication processes. C.G., K.C., S.H., and L.L. supervised the project. C.G., X.L., and L.L. wrote the manuscript.

**Data availability**

The data that support the plots within this paper and other findings of this study are available from the corresponding author upon reasonable request.

**Conflict of interest**

The authors declare no competing interests.

**Additional information**

**Supplementary information** is available for this paper.

**References:**

[1] Lu, X. et al. Emerging integrated laser technologies in the visible and short near-infrared regimes. *Nat. Photon.* **18**, 1010–1023 (2024).

[2] Labonté, L. et al. Integrated photonics for quantum communications and metrology. *PRX Quantum* **5**, 010101 (2024).

[3] Xiong, J. et al. Augmented reality and virtual reality displays: emerging technologies and future perspectives. *Light Sci. Appl.* **10**, 216 (2021).

[4] Yu, H. et al. Dual-functional triangular-shape micro-size light-emitting and detecting diode for on-chip optical communication in the deep ultraviolet band. *Laser Photonics Rev.* **18**, 2300789 (2024).

[5] Desiatov B. et al. Ultra-low-loss integrated visible photonics using thin-film lithium niobate, *Optica* 6, 380-384 (2019)

[6] Zhou, Y. et al. Common-anode LED on a Si substrate for beyond 15 Gbit/s underwater visible light communication. *Photonics Res.* **7**, 1019–1029 (2019).

[7] Karpov, M. et al. Photonic chip-based soliton frequency combs covering the biological imaging window. *Nat. Commun.* **9**, 1146 (2018).

[8] Liu, P. et al. Near-visible integrated soliton microcombs with detectable repetition rates. *Nat. Commun.* **16**, 4780 (2025).

[9] Corato-Zanarella, M. et al. Widely tunable and narrow-linewidth chip-scale lasers from near-

ultraviolet to near-infrared wavelengths. *Nat. Photonics* **17**, 157–164 (2023).

[10] Chen, H. et al. Towards fibre-like loss for photonic integration from violet to near-infrared, *Nature* **649**, 338-347 (2026)

[11] Ye, Y. et al. Visible-light optical phased array on a thin-film lithium niobate platform for high-speed beam steering. *Opt. Lett.* **50**, 3090–3093 (2025).

[12] Wu, Z. et al. Laser beam steering of 532 nm using a power-efficient focal plane array. *Opt. Lett.* **48**, 6400–6403 (2023).

[13] Tran, M. A. et al. Extending the spectrum of fully integrated photonics to submicrometre wavelengths. *Nature* 610, 54–60 (2022).

[14] He, M. et al. High-performance hybrid silicon and lithium niobate Mach–Zehnder modulators for 100 Gbit $s^{-1}$ and beyond. *Nat. Photonics* **13**, 359–364 (2019).

[15] Wang, C. et al. Integrated lithium niobate electro-optic modulators operating at CMOS-compatible voltages. *Nature* **562**, 101–104 (2018).

[16] Li, C. et al. High modulation efficiency and large bandwidth thin-film lithium niobate modulator for visible light. *Opt. Express* **30**, 36394–36402 (2022).

[17] Renaud, D. et al. Sub-1 Volt and high-bandwidth visible to near-infrared electro-optic modulators. *Nat. Commun.* **14**, 1496 (2023).

[18] Xue S. et al. Full-spectrum visible electro-optic modulator, *Optica* 10, 125-126 (2023)

[19] Villarroel, J. et al. Analysis of photorefractive optical damage in lithium niobate: application to planar waveguides. *Opt. Express* **18**, 20852–20861 (2010).

[20] Xu, Y. et al. Mitigating photorefractive effect in thin-film lithium niobate microring resonators. *Opt. Express* **29**, 5497–5504 (2021).

[21] Zhang, J., Wang, C., Denney, C. et al. Ultrabroadband integrated electro-optic frequency comb in lithium tantalate. *Nature* **637**, 1096–1103 (2025).

[22] Yu, J. et al. Tunable and stable micro-ring resonator based on thin-film lithium tantalate. *APL Photonics* **9**, 036115 (2024).

[23] Wang, H. et al. Thin-film lithium tantalate modulator operating at high optical power. *ACS Photonics* **12**, 5345–5351 (2025).

[24] Wang, C. et al. Lithium tantalate photonic integrated circuits for volume manufacturing. *Nature* **629**, 784–790 (2024).

[25] Xing, X. et al. High efficiency grating coupler on thin-film lithium niobate for visible light. *Opt. Lett.* **50**, 3361–3364 (2025).

[26] Chen, G. et al. High-performance thin-film lithium niobate modulator on a silicon substrate using periodic capacitively loaded traveling-wave electrode. *APL Photonics* **7**, 026103 (2022).

[27] Chen, X. et al. 56-m/3.31-Gbps underwater wireless optical communication employing Nyquist single carrier frequency domain equalization with noise prediction. *Opt. Express* **28**, 23784-23795 (2020).

[28] Zhou, H., Zhang, M., Wang, X., and Ren, X. Design and Implementation of More Than 50m Real-Time Underwater Wireless Optical Communication System. *J. Lightwave Technol.* **40**, 3654-3668 (2022).

[29] Sakamoto, T., Kawanishi, T. & Izutsu, M. Asymptotic formalism for ultraflat optical frequency comb generation using a Mach–Zehnder modulator. *Opt. Lett.* **32**, 1515–1517 (2007).

[30] Luo, Z. et al. 113 Gbps rainbow visible light laser communication system based on 10λ laser WDM emitting module in fiber-free space-fiber link. *Opt. Express* **32**, 2561–2573 (2024).

[31] Chen, C. et al. 100 Gbps indoor access and 4.8 Gbps outdoor point-to-point LiFi transmission systems using laser-based light sources. *J. Lightwave Technol.* **42**, 4146–4157 (2024).

[32] Hu, J. et al. 46.4 Gbps visible light communication system utilizing a compact tricolor laser transmitter. *Opt. Express* **30**, 4365–4373 (2022).

[33] Tang, L. et al. Over 23.43 Gbps visible light communication system based on 9 V integrated RGBP LED modules. *Opt. Commun.* **534**, 129317 (2023).

[34] Wang, J. et al. High-speed GaN-based laser diode with modulation bandwidth exceeding 5 GHz for 20 Gbps visible light communication. *Photonics Res.* **12**, 1186–1193 (2024).

[35] Tsai, W. S. et al. 500 Gb/s PAM4 FSO-UWOC convergent system with a R/G/B five-wavelength polarization-multiplexing scheme. *IEEE Access* **8**, 16913–16921 (2020).

[36] Wu, T. C. et al. Blue laser diode enables underwater communication at 12.4 Gbps. *Sci. Rep.* **7**, 40480 (2017).

[37] Zhou, Y. et al. Beyond 600 Gbps optical interconnect utilizing wavelength division multiplexed visible light laser communication. *Chin. Opt. Lett.* **23**, 050002 (2025).

[38] Corato-Zanarella, M., Ji, X., Mohanty, A. & Lipson, M. Absorption and scattering limits of silicon nitride integrated photonics in the visible spectrum. *Opt. Express* **32**, 5718–5728 (2024).